# BUDA-SAGE with self-supervised denoising enables fast, distortion-free, high-resolution T$_2$, T$_2^*$, para- and dia-magnetic susceptibility mapping

Zijing Zhang[1,2], Long Wang[3], Jaejin Cho[2,4], Congyu Liao[5], Hyeong-Geol Shin[6], Xiaozhi Cao[5], Jongho Lee[6], Jinmin Xu[1,2], Tao Zhang[3], Huihui Ye[1], Kawin Setsompop[5], Huafeng Liu[1 *], and Berkin Bilgic[2,4,7]

[1] State Key Laboratory of Modern Optical Instrumentation, College of Optical Science and Engineering, Zhejiang University, Hangzhou, Zhejiang, China

[2] Athinoula A. Martinos Center for Biomedical Imaging, Massachusetts General Hospital, Charlestown, MA, USA

[3] Subtle Medical Inc, Menlo Park, CA, USA

[4] Department of Radiology, Harvard Medical School, Charlestown, MA, USA

[5] Radiological Sciences Laboratory, Stanford University, Stanford, CA, USA

[6] Laboratory for Imaging Science and Technology (LIST), Department of Electrical and Computer Engineering, Seoul National University, Seoul, Republic of Korea

[7] Harvard-MIT Health Sciences and Technology, Massachusetts Institute of Technology, Cambridge, MA, United States

* Correspondence:

Huafeng Liu, PhD, liuhf@zju.edu.cn, State Key Laboratory of Modern Optical Instrumentation, College of Optical Science and Engineering, Zhejiang University, Hangzhou, 310027, ChinaRunning title: BUDA-SAGE with MR-S2S

## Abstract

**Purpose**: To rapidly obtain high resolution $T_2$, $T_2^*$ and quantitative susceptibility mapping (QSM) source separation maps with whole-brain coverage and high geometric fidelity.

**Methods**: We propose Blip Up-Down Acquisition for Spin And Gradient Echo imaging (BUDA-SAGE), an efficient echo-planar imaging (EPI) sequence for quantitative mapping. The acquisition includes multiple $T_2^*$-, $T_2'$- and $T_2$-weighted contrasts. We alternate the phase-encoding polarities across the interleaved shots in this multi-shot navigator-free acquisition. A field map estimated from interim reconstructions was incorporated into the joint multi-shot EPI reconstruction with a structured low rank constraint (BUDA reconstruction) to eliminate geometric distortion. A self-supervised MR-Self2Self (MR-S2S) neural network (NN) was utilized to perform denoising after BUDA reconstruction to boost SNR. Employing Slider encoding allowed us to reach 1 mm isotropic resolution by performing super-resolution reconstruction on BUDA-SAGE volumes acquired with 2 mm slice thickness. Quantitative $T_2$ (=$1/R_2$) and $T_2^*$ (=$1/R_2^*$) maps were obtained using Bloch dictionary matching on the reconstructed echoes. QSM was estimated using nonlinear dipole inversion (NDI) on the gradient echoes. Starting from the estimated $R_2$ and $R_2^*$ maps, $R_2'$ information was derived and used in source separation QSM reconstruction, which provided additional para- and dia-magnetic susceptibility maps.

**Results**: In vivo results demonstrate the ability of BUDA-SAGE to provide whole-brain, distortion-free, high-resolution multi-contrast images and quantitative $T_2$ and $T_2^*$ maps, as well as yielding para- and dia-magnetic susceptibility maps. Derived quantitative maps showed comparable values to conventional mapping methods in phantom and in vivo measurements.

**Conclusion**: BUDA-SAGE acquisition with self-supervised denoising and Slider encoding enables rapid, distortion-free, whole-brain $T_2$, $T_2^*$ mapping at 1 mm isotropic resolution in 90 seconds.

# Introduction

Quantitative parameter mapping has demonstrated potential in clinical and neuroscience applications (1–3), but its adoption has been hampered by the long acquisition time required to encode multi-contrast images that capture the signal evolution. In conventional mapping approaches, spin-echo (SE) acquisitions at multiple TEs (4) and multi-echo gradient echo (GRE) imaging are usually employed to obtain gold standard $T_2$ and $T_2^*$ maps. However, they generally involve long acquisition times in which only one tissue property is probed at a time, which limits their practical utility in research and clinical applications. Magnetic resonance fingerprinting (MRF) is a promising approach that can quantify multiple parameters in one scan, but still takes up to ~12 min to provide both $T_1$ and $T_2$ maps at high isotropic resolution (5–9).

Employing echo planar imaging (EPI) readout can improve the acquisition speed, but standard 2d-EPI does not lend itself to high resolution imaging due to severe geometric distortions and voxel pile-ups which stem from magnetic field inhomogeneity. EPI data also suffer from a significant resolution loss due to $T_2^*$- or $T_2$-relaxation during the lengthy readout. These artifacts have precluded EPI's ability to provide high-resolution structural images. Parallel imaging (10–13) allows for up to R=4 acceleration to reduce the readout duration, thereby mitigating, but not eliminating these artifacts. Multi-shot EPI (msEPI) encodes information from each slice across multiple acquisitions, which are then combined to provide a final image. This allows for acceleration rates of R≥8 (14,15), but requires several shots of data to provide adequate image quality at such high undersampling factors. In addition to reducing the acquisition efficiency, msEPI also fails to eliminate geometric distortions.

Here, we propose BUDA-SAGE (16–18) (Blip Up-Down Acquisition - Spin And Gradient Echo) to obtain distortion-free and high-quality multi-parameter quantitative mapping. We first add two additional EPI readouts before and after the $180^0$ refocusing pulse in a spin-echo EPI sequence to acquire gradient-, mixed-, and spin-echo images simultaneously in one scan. To provide these additional image contrasts at high in-plane resolution with high geometric fidelity, we also use an msEPI acquisition. Differently from conventional multi-shot techniques, we acquire msEPI data while alternating the phase-encoding polarity across the shots. This "Blip Up-Down Acquisition (BUDA)" technique (16,19) allows us to incorporate field inhomogeneity information into the image reconstruction, thereby eliminating geometric distortion (20). BUDA also jointly reconstructs these shots using Hankel structured low-rank regularization (14,21–23), thus obviating the need for phase navigation. By changing the TEs (echo times) of the proposed sequence, the resulting multi-contrast images can lend themselves to computation of parameter maps with high spatial fidelity, including $T_2$, $T_2^*$, para- and dia-magnetic susceptibility maps.

Although we increase the acquisition efficiency using simultaneous multislice (SMS) encoding (24–26), reaching high spatial resolution e.g. 1×1×2 mm$^3$ within a short acquisition timeframe is challenging due to SNR (signal to noise ratio) limits. To push this barrier, we employ denoising after image reconstruction. State of the art filtering approaches such as block matching and 4D filtering (BM4D) (27) and adaptive optimized non-local means (AONLM) (28) provide effective denoising ability, albeit often at the cost of over-smoothing and loss of fine scale features (29). Supervised convolutional neural networks (CNNs) have emerged as alternative and powerful techniques. Even a simple architecture comprised of convolutional kernels and rectified linear units (ReLU) such as DnCNN (30) can outperform standard filtering approaches. Supervised training involves backpropagating the network weights to minimize a loss function defined over input noisy and target clean image pairs. Challenges with a such training paradigm include the difficulty of obtaining a large of set training pairs, especially the clean images, which will likely require an increased number of signal averages or shots. Supervised networks may also suffer from poor generalizability e.g. when acquisition parameters such as resolution, TE or repetition time (TR) are different in the test dataset. To gain robustness to such issues, we propose to use a self-supervised neural network for denoising. Self-supervised deep-learning denoising methods have shown great potential in improving image SNR (31–33), and can complement rapid EPI acquisitions. Inspired by Self2Self (S2S) (33), a network trained in a self-supervised manner for natural image denoising, we build a deep model, called MR-S2S, which is trained using only noisy images without ground truth information. We further show that MR-S2S provides comparable denoising performance to a supervised model.

gSlider (generalized Slice dithered enhanced resolution) is a radiofrequency (RF) encoding technique for performing SNR efficient volumetric simultaneous multi-slab acquisitions (34). To achieve high resolution in the slice direction, a slab is volumetrically encoded using RF pulses with a scheme similar to Hadamard encoding. Inspired by gSlider, we propose to achieve 1 mm isotropic resolution quantitative mapping using a simpler Slider approach (35), which utilizes sub-voxel spatial shifts in the slice direction. This enables us to perform super-resolution reconstruction for increasing both resolution and SNR efficiency synergistically with MR-S2S.

By combining BUDA-SAGE with MR-S2S and Slider, we propose to harness the efficiency of this distortion-free acquisition in high-resolution imaging with high SNR and enable fast and high-fidelity multi-parameter mapping, and demonstrate whole-brain $T_2$ and $T_2^*$ maps at 1 mm isotropic resolution in 90 seconds. We further obtain separate para- and dia-magnetic susceptibility maps from a ~140 second acquisition, thereby providing detailed information on tissue composition that may facilitate neurodegeneration and aging research. Phantom and in vivo results show that the $T_2$ and $T_2^*$ maps estimated by the proposed method are consistent with the slower gold standard mapping acquisitions. BUDA-SAGE reduces the scan time dramatically (more than 20× compared to gold standard methods) for increased motion robustness, while providing clinically desired, multi-contrast data as well as multi-parameter maps.

## Methods

### BUDA-SAGE acquisition

In spin echo EPI acquisitions with late TEs required for $T_2$ mapping, there are two potential periods with unused sequence time in each TR: between the $90^0$ excitation pulse and the $180^0$ refocusing pulse, and between the refocusing pulse and the spin echo readout. We propose to exploit this unused sequence time by inserting two additional EPI readouts, thereby collecting three different contrasts ($T_2^*$-, $T_2'$- and $T_2$-weighted images with corresponding TEs, $TE_{gre}$, $TE_{mixed}$, $TE_{se}$) in one scan. Figure 1(a) shows the proposed BUDA-SAGE sequence, where we have also employed SMS encoding with blipped-CAIPI acquisition (25) to further boost the efficiency.

In order to mitigate distortion and blurring and achieve high resolution while enabling a wider range of TEs using high in-plane acceleration, a multi-shot EPI acquisition was incorporated. Alternating shots were sampled with opposite phase-encoding polarities (blip up and down).

**BUDA-SAGE reconstruction**

With the acquired interleaved blip-up and -down shots, distortion-free images of each contrast and each scan can be reconstructed using the pipeline shown in Figure 1(b), which includes the following:

1. For each contrast, the k-space data for the blip-up shots and blip-down shots are summed in sliding-window fashion, and SENSE reconstruction is performed to generate interim blip-up and blip-down image pairs. Stemming from field inhomogeneity, the SENSE blip-up/down images have opposite geometric distortions due to reversed phase-encoding direction (voxel pile-ups can also be seen as denoted by the red arrows in Figure 1(b)).
2. All of the interim blip-up/down image pairs are jointly processed in FSL *topup* (http://fsl.fmrib.ox.ac.uk/fsl) to estimate a field map (36).
3. The estimated field map is incorporated into the Hankel structured low-rank constrained joint reconstruction. This can be described as:

$$\min_x \sum_{s=1}^{Ns} \|F_s W_s C x_s - d_s\|_2^2 + \lambda \|H(x)\|_* \quad (1)$$

Where $F_s$ is the under-sampled Fourier operator in shot $s$, $W_s$ is distortion operator (off-resonance information estimated from all interim blip-up/down image pairs) in shot $s$, $C$ is the coil sensitivity map estimated from distortion-free spin-warp gradient-echo calibration data using ESPIRIT (12), and $d_s$ is the k-space data of each shot. $\|H(x)\|_*$ enforces low-rank prior on the block-Hankel representation of the blip-up and blip-down data, which is implemented by selecting n×n (n=7 in all reported results) neighborhood points in k-space from each shot and then concatenating them in the column dimension. The reconstruction is implemented using a projection onto convex sets (POCS) like (37,38) iteration that alternates between data consistency and low-rank truncation steps with the tolerance of 0.01% RMSE between two successive iterations.

Note that the three pairs of blip-up/down interim SENSE images ($T_2^*$, $T_2'$ and $T_2$ weighted) from each scan are all provided as inputs to FSL to jointly estimate a field map, rather than using only one pair of images based on the ability of the SAGE sequence to provide multiple contrasts.

**Self-supervised Denoising with MR-S2S**

Taking noisy images as inputs, MR-S2S employs convolutions and deconvolutions at the beginning and end of the model architecture for image compression and upsampling, respectively. In the middle part of the MR-S2S, multiple ResBlocks (33) are constructed to extract hierarchical features from the image. Each ResBlock includes stacked convolutional filters paired with ReLU activation functions and dropout regularization. Model architecture is provided in Figure 2. To train the proposed MR-S2S, we use a mask **m** whose elements are independent identically distributed (*iid*) sampled from a Bernoulli distribution with parameter $\mu$ (0.3 in our experiments) at each training epoch. To form the input data of the network, the mask **m** is voxel-wise multiplied with the noisy images (all contrasts along the echo time *t* to form image volumes which are obtained by BUDA-SAGE reconstruction). At the same time, we obtain the corresponding complementary mask, 1-**m**, which is also voxel-wise multiplied with the same noisy image to provide the target image. We minimize the L2-norm loss between the outputs (the results after going through MR-S2S network) and targets to train the MR-S2S network. A single MR-S2S model was trained for each slice of all contrasts.

In the test phase, for each noisy image, we independently sample multiple Bernoulli masks (we used 100 in the experiments) to obtain masked images; at the end of MR-S2S, we average the outputs from the corresponding masked images as the final result, leading to effective image denoising.

The MR-S2S network was implemented using Keras application programming interface (API) with a Tensorflow backend. The training was performed using a V100 GPU (NVIDIA, Santa Clara, CA). The L2-norm loss was used to optimize the NN's parameters using the Adam optimizer (39) with default parameters (except for the learning rate). The setting details of the parameters are as follows: number of ResBlocks = 10, number of filters = 64, learning rate = 0.0005, number of epochs = 30, kernel size = 3×3, batch size = 4.

**Slider super-resolution reconstruction**

Slider acquires $N_{slider}$ sets of high SNR, thick slice data (e.g. at resolution 1×1×$N_{slider}$ mm$^3$) with 1 mm shift between each set along the slice direction to achieve 1 mm isotropic resolution final results. The $N_{slider}$ =2 interleaved acquisition used in our experiments is displayed on Figure 10(a). To achieve thin-slice high-resolution reconstruction, data are first reconstructed using BUDA-SAGE reconstruction with MR-S2S denoising. The thin slices are then resolved within the acquired thick slice images using linear reconstruction with Tikhonov regularization:

$$z = (A^T A + \lambda_{Tik} I)^{-1} A^T b = A_{inv\_Tik} b \qquad (2)$$

Where *A* is the Slider encoding matrix that sums adjacent slices, *b* is the concatenation of the acquired Slider-encoded thick slice images (at a given x, y position), and *z* is the corresponding high slice resolution reconstruction. $\lambda_{Tik}$ is the Tikhonov regularization parameter and was set to 0.2 to achieve a high SNR gain and sharp slice resolution based on a retrospective experiment, which will be shown in Supporting information Figure S3. We note that the *A* matrix employed in our reconstructions did not include slice profile information, but this could be flexibly incorporated into Eq2.

**Quantitative mapping**

In BUDA-SAGE EPI acquisition, signal evolution obeys the following equation (17):

$$S(t) = \begin{cases} S_0^I \cdot e^{-t \cdot R_2^*} & , 0 < t < TE_{SE} \\ S_0^{II} \cdot e^{-TE_{SE} \cdot (R_2^* - R_2)} \cdot e^{-t \cdot (2 \cdot R_2 - R_2^*)} & , TE_{SE}/2 < t < TE_{SE} \end{cases} \quad (2)$$

Where $S(t)$ is the MRI signal acquired at the echo time $t$ in a BUDA-SAGE EPI time series. Voxel-wise estimates of $S_0^I$, $S_0^{II}$, $R_2$ (=$1/T_2$) and $R_2^*$ (=$1/T_2^*$) are obtained through a least-squares solution of equation (2) from all signal values acquired in our experiments. From these estimates, an additional parameter $\delta = S_0^I / S_0^{II}$ is obtained, which accounts for residual signal difference owing to imperfectly matched slice profiles between echo trains acquired before and after the refocusing $180^0$ pulse (40). The voxel-wise estimates of $\delta$ are assumed to be constant for the whole BUDA-SAGE EPI acquisition. Therefore, the least-squares solution of equation (2) for each time point consisted of three instead of four unknown values, namely, $\delta$, $R_2$ and $R_2^*$. With this three-parameter fit, it is possible to achieve better temporal signal stability compared to a four-parameter fit (17).

The values of $\delta$ in each voxel estimated from the three-parameter fit are used in a subsequent two-parameter Bloch dictionary match. We discretize the three parameters, $\delta$, $R_2$ and $R_2^*$, to build a dictionary adapted to the BUDA-SAGE data. The range of $\delta$ was determined by the values estimated from three-parameter iteration fit to be in 1.00-1.65, which we discretized to one hundred values to form the dictionary. For each $\delta$, we use discretized $T_2$ and $T_2^*$ values to form the signal $S$ based on equation (2). We set the $T_2$ values as [1:1:50; 52:2:150; 155:5:500]; the $T_2^*$ values as [1:1:50; 52:2:150; 155:5:300]. We match the signal $S$ (all the echoes after BUDA-SAGE reconstruction and MR-S2S denoising) with the dictionary for each voxel by maximizing their dot product, which yields the corresponding $T_2$ and $T_2^*$ values.

$R_2$, $R_2^*$ and $R_2'$ (= $R_2^*$-$R_2$) maps are also obtained after the dictionary match. Susceptibility maps are estimated using the gradient echoes from the acquisition with nonlinear dipole inversion (NDI) (41) which is based on the nonlinear-MEDI (NMEDI) (42) approach, but magnitude weighting and nonlinear formulation are exploited as inherent regularizers along with a simple Tikhonov penalty. Phase pre-processing steps included brain masking using FSL BET (43), Laplacian phase unwrapping (44,45), and V-SHARP (46,47) background removal with the largest kernel size of 25 voxels. Using the estimated $R_2'$ information, susceptibility source separation reconstruction is also performed, which provides additional para- and dia-magnetic susceptibility maps (48).

**Phantom Validation**

To validate the proposed method, a tube phantom with twelve ROIs was scanned. BUDA-SAGE data were acquired using the following protocol: in-plane resolution = 1×1 mm, slice thickness = 2mm, FOV = 220×220×52 mm$^3$, 26 slices, R$_{in-plane}$ = 8, SMS factor = 1. We changed the TEs of the sequence and performed three separate scans to acquire three groups data with [TE$_{gre}$, TE$_{mixed}$, TE$_{se}$]=[18, 64, 91] ms in group 1, [30, 88, 115] ms in group 2 and [42, 112, 139] ms in the last scan (group 3) at TR = 4500 ms. Each group provided three different contrasts. We collected 8-shots of BUDA-SAGE with interleaved blip-up and blip-down phase encoding to cover the k-space. The total acquisition time was 112.5s ($T_{acq}$ = TR × $N_{shot}$ × $N_{group}$ + $T_{dummy}$ = 4.5s×8×3+4.5) with a total of 24 shots ($N_{total-shot}$ = $N_{shot}$ × $N_{group}$ = 8×3) along with a dummy shot to enter steady-state.

The quantitative maps from the proposed acquisition were compared to maps obtained using conventional mapping methods. To estimate the T$_2$* map, a 2D multi-echo GRE sequence with 10 different TEs (from 6 ms to 60 ms with a gap of 6ms), TR = 4.5s, resolution = 1×1×2 mm, 26 slices and FOV = 192×192×54 mm$^3$ was acquired in a scan time of 4 min 48s. To derive a T$_2$ map, six multi-slice single spin echo acquisitions with different TEs (from 19 ms to 139 ms with a gap of 24 ms), TR = 4.5s, resolution = 1×1×2 mm, 26 slices and FOV = 192×192×54 mm$^3$ were acquired in a total scan time of 28 min 48s.

All experiments above were performed on a 3 Tesla (T) MAGNETOM Prisma scanner (Siemens Healthcare, Erlangen, Germany) with a 20-channel head receiver coil, which accommodated the large size of the phantom.

**In-vivo Validation**

To validate our proposed method on in-vivo data, nine healthy volunteers were scanned with the approval of Institutional Review Board.

BUDA-SAGE data were collected using the following protocol: in-plane resolution = 1×1 mm, slice thickness = 2 mm, FOV = 220×220×120mm$^3$, 60 slices, R$_{in-plane}$ = 8, SMS factor = 2. We changed the TEs of the sequence and performed three separate scans to acquire three groups of data with the same nine TEs as in the phantom experiment at TR = 5000 ms. Each group provided three different contrasts as well. As a reference, 8-shot BUDA-SAGE was acquired with a phase-encoding shift Δk$_y$ of 1 between each shot (interleaved blip-up and blip-down) in order to cover the whole k-space and perform parallel imaging reconstruction. The total acquisition time was 125s ($T_{acq}$ = TR × $N_{shot}$ × $N_{group}$ + $T_{dummy}$ = 5s×8×3+5) with a total of 24 shots ($N_{total-shot}$ = $N_{shot}$ × $N_{group}$ = 8×3) as well as one initial dummy shot. This was then subsampled to 4-shot (2 blip-up, 2 blip-down, seen in Figure S1) and 2-group (one of the 3 groups was omitted), resulting in 8 shots ($N_{total-shot}$ = $N_{shot}$ × $N_{group}$ = 4×2) as well as a dummy shot, corresponding to an acquisition time of

45s ($T_{acq} = TR \times N_{shot} \times N_{group} + T_{dummy} = 5s \times 4 \times 2 + 5$). Note that the phase-encoding shift $\Delta k_y$ is now 2.

To validate the accuracy of $T_2$ and $T_2^*$ maps derived from our proposed method, we acquired three groups of BUDA-SAGE data and the conventional mapping acquisitions on a second volunteer. BUDA-SAGE acquisition parameters were kept the same. For conventional mapping acquisitions, to quantify $T_2^*$ map, a multi-echo GRE sequence with 10 different TEs (from 6 ms to 60 ms with a gap of 6ms), TR = 5s, resolution = 1×1×2 mm, 60 slices and FOV = 220×206×120 mm$^3$ was acquired in a scan time of 8 min 35s. To derive a $T_2$ map, six multi-slice single spin echo acquisitions with different TEs (from 19 ms to 139 ms with a gap of 24 ms), TR = 7s, resolution = 1×1×2 mm, 60 slices and FOV = 192×174×120 mm$^3$ were acquired in a total scan time of 40 min 36s.

To obtain quantitative susceptibility map and disentangle its para- and dia-magnetic constituents, we collected another three groups of BUDA-SAGE data on a third volunteer. The acquisition parameters are same as the first two experiments except for FOV = 220×220×128mm$^3$ with 64 slices, TR = 5500 ms. Thus, the total acquisition time was 137.5s ($T_{acq} = TR \times N_{shot} \times N_{group} + T_{dummy} = 5.5s \times 8 \times 3 + 5.5$).

To conduct a supervised NN experiment and demonstrate the effectiveness of the self-supervised MR-S2S on denoising BUDA-SAGE reconstruction, we collected 2-averages (scanned the same protocol twice: the setting of parameters is same as the first two datasets) of 8-shot BUDA-SAGE data on five additional subjects. The averaged BUDA-SAGE images were used as reference information for training and validating the supervised NN model. We note that the training pairs were: 2-average 8-shot BUDA-SAGE images (ground truth, 90 sec scan for each group) and 4-shot (a subset from 8-shot data, 25 sec scan for each group) BUDA-SAGE images (noisy images).

To provide T2 and T2* maps at 1 mm isotropic resolution, two groups of BUDA-SAGE data were acquired on a ninth volunteer. The acquisition parameters were the same as the first experiment. We collected each of these two groups data at the slice thickness of 2 mm. To enable super-resolution reconstruction in the slice direction, we performed an additional acquisition by shifting the slice position by 1 mm for each group. Performing super-resolution reconstruction on this Slider encoded data yielded the final 1 mm isotropic volumes. The total acquisition time for this experiment was 85s ($T_{acq} = TR \times N_{shot} \times N_{group} \times N_{slider} + T_{dummy} = 5s \times 4 \times 2 \times 2 + 5$).

An FOV-matched low resolution 2D GRE was also acquired to obtain distortion-free sensitivity maps for all BUDA-SAGE reconstructions from a 2 sec calibration scan.

All experiments were performed on a 3T MAGNETOM Prisma scanner (Siemens Healthcare, Erlangen, Germany) with a 32-channel head receiver coil. Computations were performed on a Linux (Red Hat Enterprise) server (with Core i7 Intel Xeon 2.8 GHz CPUs and 64GB RAM) using MATLAB R2014a (The MathWorks, Inc., Natick, MA).

## Results

The distortion-free multi-contrast tube phantom reconstruction results using 8-shot BUDA-SAGE acquisition from three groups (each group contains a $T_2^*$-, a $T_2'$- and a $T_2$-weighted image) was shown in Supporting Information Figure S2. $T_2$ and $T_2^*$ maps were obtained using Bloch dictionary matching using these multi-contrast images (Figure 3(a), top). Reference $T_2$ and $T_2^*$ maps were calculated by performing a voxel-wise mono-exponential fit to the echo images (Figure 3(a), bottom). Overall, good agreement is seen between the proposed and conventional mapping methods. The quantitative maps from the BUDA-SAGE sequence were compared to the maps acquired with conventional methods from the same slice. Bland-Altman analysis was performed on the twelve ROIs inside the tube phantom to evaluate the agreement between maps, as shown in Figure 3(b). The ROIs show similar values between these two methods and all points were within the limits of agreement, though some slight biases ($T_2$: -2.1 ms, $T_2^*$: -6.6 ms) were observed.

T2 and T2* maps from in-vivo data were also compared between MB2 BUDA-SAGE and conventional acquisitions, as shown in Figure 4(a). Similarly, good agreement is observed in the displayed slice. Six ROIs were defined as 9×9 voxel boxes throughout the slice, and the results from Bland-Altman analysis are shown in Figure 4(b). These ROIs demonstrate similar values between methods and all points were within the limits of agreement, although some slight biases ($T_2$: 4.5 ms, $T_2^*$: 0.5 ms) were seen.

Applying MR-S2S after BUDA-SAGE reconstruction, we obtained distortion-free, high-resolution (1×1×2 mm$^3$) multi-contrast results including $T_2^*$-, $T_2'$- and $T_2$-weighted images using a 4-shot acquisition (subsampled from 8-shot BUDA-SAGE data), and group 1 is shown in Figure 5. As illustrated in the zoom-in, BUDA+MR-S2S reduces the noise as well as preserving fine details compared to standard BUDA reconstruction. To further validate the effectiveness of MR-S2S, we conducted an experiment with a supervised NN whose network structure is the same as MR-S2S. The difference between BUDA+MR-S2S and BUDA+supervisedNN is that the latter uses reference data as supervised information for training. As seen in Figure 5, the results of BUDA+MR-S2S are comparable to those of BUDA+supervisedNN, both qualitatively and quantitatively based on the RMSE values.

Figure 6 shows whole-brain distortion-free six-contrast (two $T_2^*$-, two $T_2'$- and two $T_2$-weighted) images in sagittal, coronal and axial views using two groups of 4-shot BUDA-SAGE data with MR-S2S denoising from a 45s acquisition (20s/group + 5s dummy scan), plus a 2s calibration scan.

Figure 7 shows whole-brain, distortion-free $T_2$ and $T_2^*$ maps at 1×1×2 mm$^3$ resolution in sagittal, coronal and axial views using the BUDA-SAGE acquisition. As a reference, we have used the maps acquired from 8-shot, 3-group data (resulting in a total of 9 contrasts: three $T_2^*$-, three $T_2'$- and three $T_2$-weighted) as shown in the first column of Figure 7. The total acquisition time for the reference scan was 127s (40s/group + 5s dummy scan + 2s calibration). The maps are obtained using Bloch matching after BUDA-SAGE reconstruction. The other three are quantitative maps using 4-shot (a subset of 8-shot), 2-group (using groups #1 and #3) data. The second column shows

$T_2$ and $T_2^*$ maps obtained without additional denoising. Third and fourth columns show parameter maps estimated from echoes with MR-S2S and supervised NN denoising, respectively. Based on the RMSE values (14.9% vs. 15.5% for $T_2$ map and 18.7% vs. 19.2% for $T_2^*$ map), denoising performance of MR-S2S was seen to be slightly better than that of supervised NN.

Figure 8 displays whole-brain, distortion-free total, paramagnetic (e.g., iron) and diamagnetic (e.g., myelin) QSM maps along with $R_2$, $R_2^*$ maps at $1\times1\times2$ mm$^3$ resolution in sagittal, coronal and axial views using 3-groups of 8-shot data. The total acquisition time is 139.5s (44s/group + 5.5s dummy scan + 2s calibration). While iron-rich basal ganglia nuclei were well localized in the $\chi_{positive}$ map, white matter structures such as the optic nerve were visible in the $\chi_{negative}$ component.

Figure 9(b) shows the whole-brain, distortion-free, ~1 mm isotropic resolution Slider super-resolution reconstruction $T_2^*$-weighted images in the sagittal and coronal views. These volumes are obtained after applying BUDA-SAGE reconstruction and MR-S2S denoising on two $1\times1\times2$ mm$^3$ resolution images (each is a 4-shot acquisition which is subsampled from 8-shot data) with 1 mm slice position shift in the BUDA-SAGE acquisition (namely, $N_{slider} = 2$ interleaved acquisition). Similarly, the $T_2$'- and $T_2$-weighted ~1 mm isotropic images are displayed on Figure 9(c), demonstrating the ability of Slider reconstruction to provide high-resolution data.

Supporting Information Figure S4 displays $T_2^*$-weighted images at $1\times1\times$~1 mm$^3$ resolution obtained using a range of $\lambda_{Tik}$ (here we show three values) values for Tikhonov regularization of Slider super-resolution reconstruction. The impulse responses of the central slice within an Slider-encoded thick slice were shown in the bottom. Here, $b$ was replaced with a Slider-encoded delta function of the central slice and passed through the reconstruction method to obtain the impulse response (19). The full-width-half-max (fwhm) resolution in the slice direction ($\Delta z_{eff}$) were also calculated from the impulse response. The reference results were obtained using 8-shot acquisition with $\lambda_{Tik}$ of 0.1. Judging from RMSE and $\Delta z_{eff}$ values, $\lambda_{Tik}$ was selected 0.2 as a trade off to achieve a high SNR gain and sharp partition resolution. Though we can get higher $\Delta z_{eff}$ for $\lambda_{Tik} = 0.02$, the banding artifacts and noise amplification are obvious from the images. And clear blurring in partition direction can be found for $\lambda_{Tik}$ of 1.

Figure 10 shows whole-brain, distortion-free $T_2$ and $T_2^*$ maps at $1\times1\times$~1 mm$^3$ resolution in sagittal, coronal and axial views using two groups of 4-shot Slider-encoded ($N_{slider} = 2$) BUDA-SAGE acquisition from a 85s scan (20s/group/slider + 5s dummy scan), plus a 2s calibration scan for coil sensitivities.

## Discussion

In this work, we have demonstrated rapid, distortion-free, multi-contrast and quantitative imaging using a synergistic combination of efficient multi-shot SAGE acquisition, joint BUDA reconstruction, Slider encoding and self-supervised MR-S2S denoising. This technique took advantage of the following:

1. Multi-contrast EPI: two readouts were inserted before and after $180^0$ refocusing pulse in a spin-echo EPI sequence to collect additional $T_2^*$- and $T_2'$-weighted contrasts in one scan. By changing the TEs, additional contrasts were obtained rapidly.

2. Distortion-free EPI: alternating the phase-encoding polarities across the shots during a multi-shot acquisition allowed for eliminating geometric distortion using our BUDA technique.

3. SNR boost using self-supervised denoising: we have adopted MR-S2S denoising to improve the SNR of rapidly acquired BUDA-SAGE data. This network was trained without an external reference dataset, while performing similarly as its supervised counterpart. Such self-supervised training paradigm helps increase robustness against potential generalization issues supervised networks may suffer from, and provides an SNR boost to the multi-contrast images.

4. Slider encoding: by performing multiple thick-slice acquisitions and super-resolution reconstruction, image SNR was further boosted to permit high isotropic resolution imaging using EPI readouts.

These technical developments permitted fast and high-resolution multi-parameter mapping, through which whole-brain, $1\times1\times2\text{mm}^3$ $T_2$ and $T_2^*$ maps were obtained by Bloch dictionary matching using distortion-free, high-SNR, multi-contrast images from a 47s scan. Incorporating Slider encoding further allowed for parameter mapping at 1 mm isotropic resolution from a 87s acquisition. The accuracy of our $T_2$ and $T_2^*$ maps were verified both on phantom and in-vivo data compared to conventional mapping methods. Total QSM and additional para- and dia-magnetic susceptibility maps were estimated from a ~140s scan, which could be further sped-up less number of shots and MR-S2S denoising.

We hope that such rapid quantitative acquisitions will contribute to our community's efforts towards improving the value of MRI. An important clinical utility will be in scanning vulnerable patient populations such as pediatric and elderly subjects, who may have difficulty remaining still during lengthy data acquisitions. BUDA-SAGE with MR-S2S reduces the scan period dramatically for increased motion robustness, while providing clinically desired, multi-contrast data and multi-parametric maps. We anticipate further applications in scanning emergent or in-patient populations, where the subjects are often acutely ill and have difficulty complying with long scans.

The quantitative maps estimated from this multi-contrast EPI approach showed good agreement with conventional mapping techniques. The $T_2$ and $T_2^*$ maps from the BUDA-SAGE acquisition showed slightly different values in some regions compared to maps from the conventional acquisitions. The maps estimated from our multi-contrast EPI approach has smaller number of echoes contributing to the $T_2/T_2^*$ estimation. Nevertheless, the maps both from phantom and in-vivo data gave consistent results, and in particular, the SMS acquisition did not influence the results. Recently, a similar study from (49) also shows the observed biases in T2 and T2* maps. Our approach was able to provide better agreement with conventional mapping methods as well as achieving much thinner slice resolutions with whole brain coverage.

Growing interest in QSM's ability to disentangle para- and dia-magnetic contributions to tissue susceptibility is fueled by emerging applications in imaging neurodegenerative diseases. Having the ability to disentangle iron from myelin or other diamagnetic proteins (e.g. amyloid and tau) will contribute to our understanding of disease mechanisms in e.g. Multiple Sclerosis (MS) (50) and Alzheimer's Disease (AD) (51). Source separation QSM is an emerging approach that could help accomplish this goal, along with other influential modeling approaches (45,52). With BUDA-SAGE, we have taken a step towards accelerating the demanding data acquisition aspect of such models, where additional quantitative maps (e.g. $R_2^*$, $R_2'$) are often needed to fit model parameters.

We think that the ability of self-supervised training to perform denoising without clean targets stems from two factors. First, from a statistical perspective, the expectation of our training loss is the same as that we use noisy input to predict the ground truth even though there is no ground truth on condition that we assume the noise components are independent and of zero mean. In addition, the noise variance will dramatically increase when the number of training samples decreases from many to one from a Bayes estimator view. Thus, variance reduction is the key for the self-supervised learning on so few images. Therefore, second, we adopt a data augmentation approach which samples complementary Bernoulli masks randomly to generate many pairs of training samples to prevent the network from converging to an identity map. A dropout step is employed to randomly prune nodes to increase the statistical independence of our outputs, and thus the average of these predictions will reduce the variance of the result.

For comparison, the reconstruction results were denoised using the state-of-the-art BM4D (27) denoising algorithm, an extension of the BM3D algorithm for volumetric data. Briefly, the BM4D method groups similar 3-dimensional blocks into 4-dimensional data arrays to enhance the data sparsity and then performs collaborative filtering to achieve superior denoising performance. The BM4D denoising was performed assuming Rician noise with an unknown noise standard deviation on the magnitude data and was set to estimate the noise standard deviation and perform collaborative Wiener filtering with "modified profile" option. From the zoom in part of Figure S3, it can be seen that MR-S2S denoises the data while better preserving the image details. However, the BM4D result appears to suffer from over-smoothing. MR-S2S also behaves better than BM4D quantitatively seen from RMSE. In terms of denoising speed, BM4D (~ 17.5 sec whole volume) is faster than self-supervised MR-S2S (~ 2.6 min/slice which depends on the number of epochs).

**Limitations and Potential Extensions:** The relatively narrow range of the acquired TEs may be a limitation in the application of the proposed technique. To help address this drawback, we have chosen the high in-plane acceleration rate of R = 8 to largely mitigate $T_2/T_2^*$ blurring as well as making the minimum TE as short as possible. In our current acquisitions, the minimum TEs of $T_2^*$, $T_2'$ and $T_2$ weighted echoes are 18 ms, 64 ms and 91 ms, which are constrained by the readout duration. The latest TEs were chosen to be 42 ms, 112 ms and 139 ms. Though these values are used to target gray and white matter in the brain, this range may not lend itself to imaging short $T_2$ and $T_2^*$ species (e.g. cartilage) or very long species (e.g. CSF). Further reductions in TE are possible through partial Fourier, where BUDA acquisition could provide complementary information by sampling the opposite sides of k-space through phase encoding polarity reversal.

There are remaining artifacts (seen from the red arrow in Figure 8) stemming from intra-voxel $B_0$ field inhomogeneity effects in the estimated $R_2^*$ maps. These can be mitigated by employing sinc functions (53) arising from Fourier transform of the voxel basis functions modulated by intra-voxel linear phases in the dictionary matching model, but this will also increase noise level in the relaxation results. Employing regularization during V-SHARP filtering (54) could help dampen the impact of these artifacts on tissue phase and susceptibility maps.

To speed-up MR-S2S, it can be possible to pre-train the model on a set of subjects, then fine-tune it on each specific test subject's data using much fewer number of epochs. Extending MR-S2S to perform volumetric denoising could provide additional performance gain, and make this approach more amenable to fine-tuning since there will be a single model trained for the entire volume, rather than a different network per slice. In the part of image reconstruction, the time depends on several factors, including the number of central processing unit (CPU) cores used for parallel imaging, image size, number of shots, number of iterations, and stopping criteria, but should be able to be sped up using several multi-shot reconstruction methods (55,56) proposed recently.

## Conclusion

In this work, BUDA-SAGE with self-supervised denoising was proposed as a new approach for distortion-free multi-contrast and quantitative imaging with the ability to provide whole-brain coverage at $1\times1\times2m^3$ resolution under 50 seconds, and using Slider encoding, 1 mm$^3$ isotropic resolution under 1.5 minutes. This provides a favorable balance between acquisition speed, image quality, and the accuracy of $T_2$ and $T_2^*$ estimation. Furthermore, our distortion-free, multi-shot and multi-echo acquisition also enables fast, high-resolution para- and dia-magnetic susceptibility mapping based on source separation QSM reconstruction.

## Acknowledgement

This work was supported in part by the National Natural Science Foundation of China (No: U1809204, 61525106, 61427807, 61701436), by the National Key Technology Research and Development Program of China (No: 2017YFE0104000, 2016YFC1300302), by Shenzhen Innovation Funding (No: JCYJ20170818164343304, JCYJ20170816172431715),
and by NIH R01 EB028797, R03 EB031175, U01 EB025162, P41 EB030006, U01 EB026996 and the NVidia Corporation for computing support.
Zijing Zhang was supported by the China Scholarship Council for 2 years at Massachusetts General Hospital.

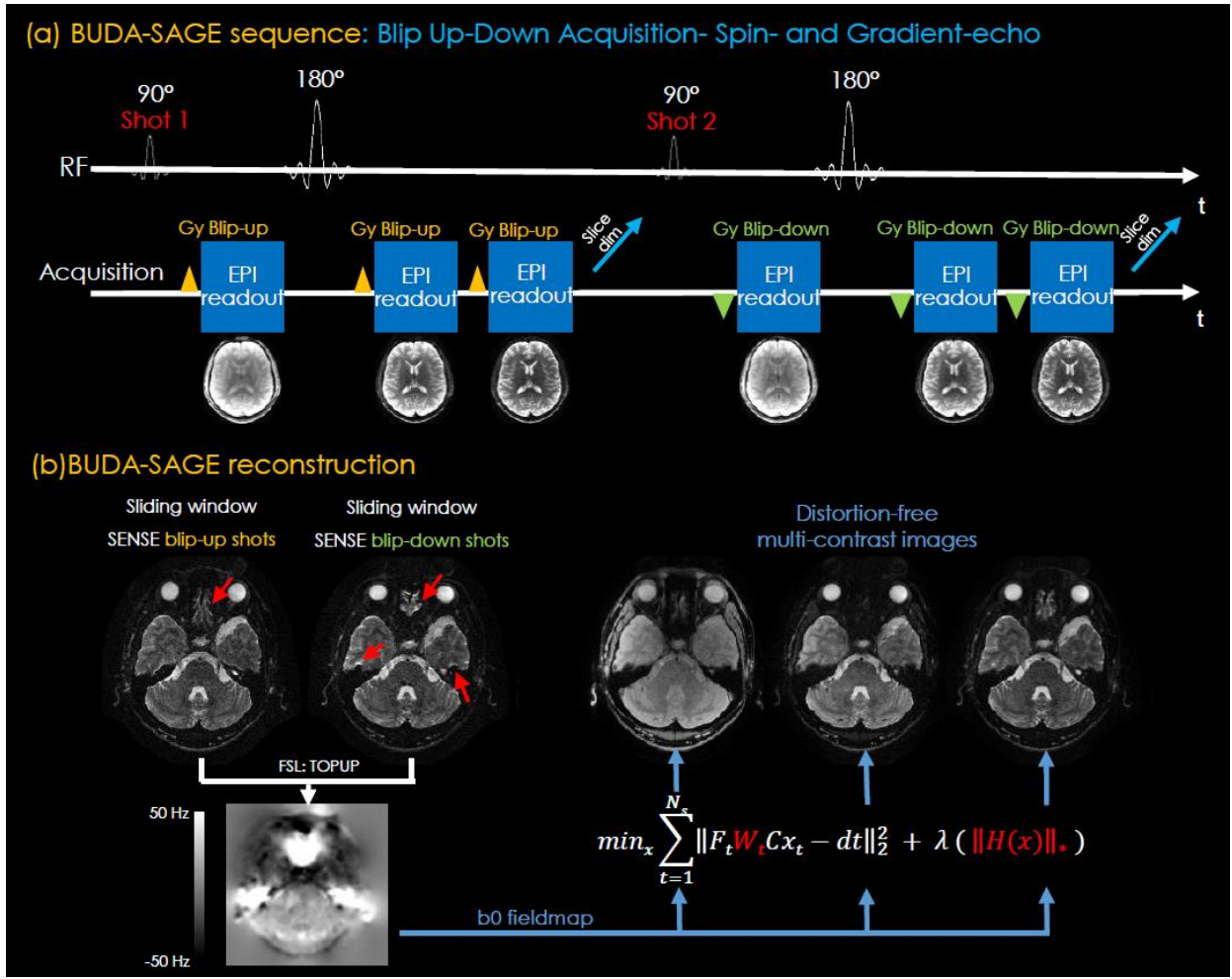

Figure 1

(a) The sequence diagram of the BUDA-SAGE (blip up-down acquisition- spin- and gradient-echo) acquisition. Each scan can provide 3-echoes (a gradient echo, a mixed gradient-and spin echo, and a spin echo). Multiple scans can provide additional contrasts by changing the TEs of the sequence.

(b) The pipeline of BUDA-SAGE reconstruction. We conduct sliding window SENSE reconstruction for blip-up and blip-down shots, then use them jointly in FSL *topup* to estimate a B0 field map. We incorporate this B0 field map into the Hankel structured low-rank constrained parallel imaging to jointly reconstruct distortion-free images for all echoes.

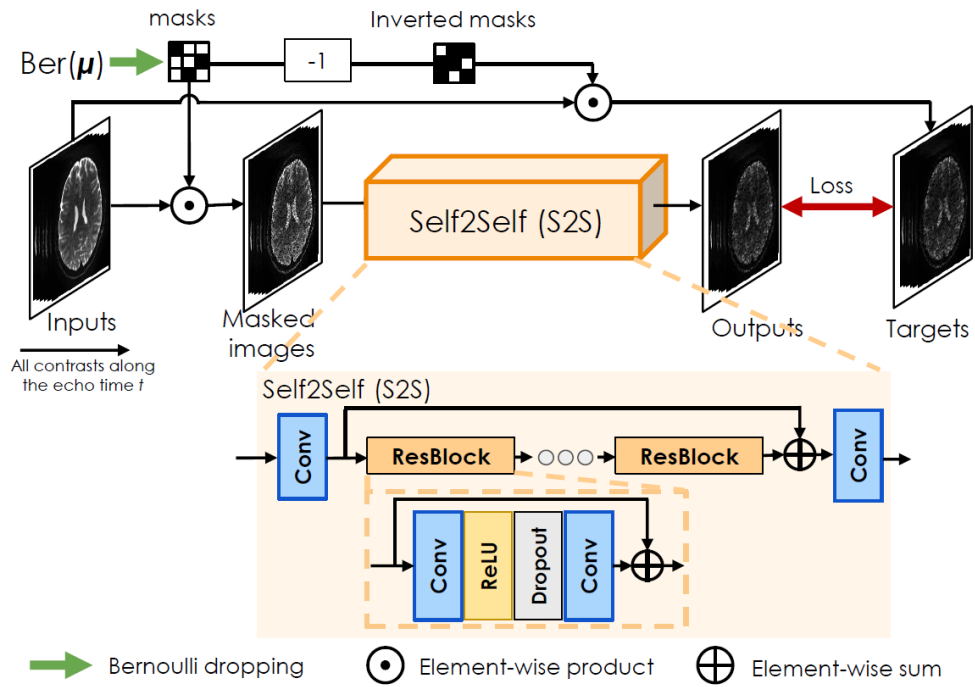

Figure 2

Network diagram for the self-supervised MR-Self2Self (MR-S2S) model for image denoising after BUDA-SAGE reconstruction.

MR-S2S takes noisy images as inputs and applies masks whose elements are *iid* sampled from a Bernoulli distribution to implement a dropout layer in the beginning. The corresponding inverted masks are applied on the noisy inputs to form the target images of MR-S2S.

## (a) Quantitative maps comparison

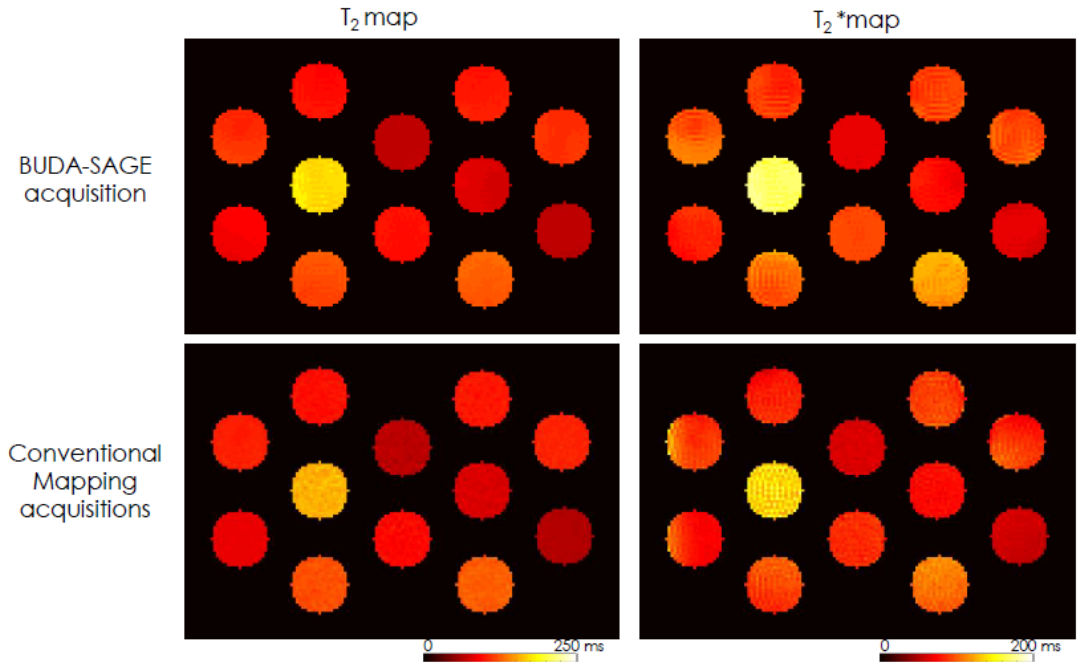

## (b) Bland-Altman plots

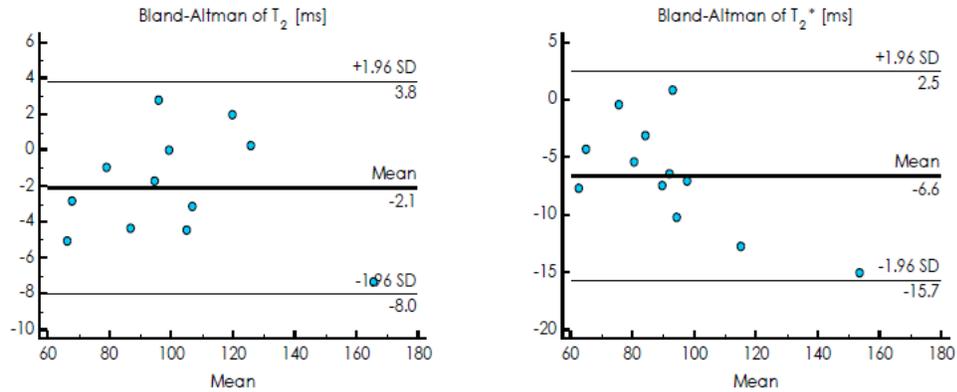

Figure 3

(a) Distortion-free quantitative $T_2$ and $T_2^*$ maps of a tube phantom at $1 \times 1 \times 2$ mm$^3$ resolution were obtained using Bloch dictionary matching using the multi-contrast images. The maps from a BUDA-SAGE msEPI acquisition (top) are compared with maps derived from conventional approaches (bottom). Twelve ROIs from the slice correspond to the twelve tubes present in the phantom.

(b) The Bland-Altman plots showing the mean and difference for the ROIs for $T_2$ and $T_2^*$ values. All values are within the limits of agreement though minor biases remain.

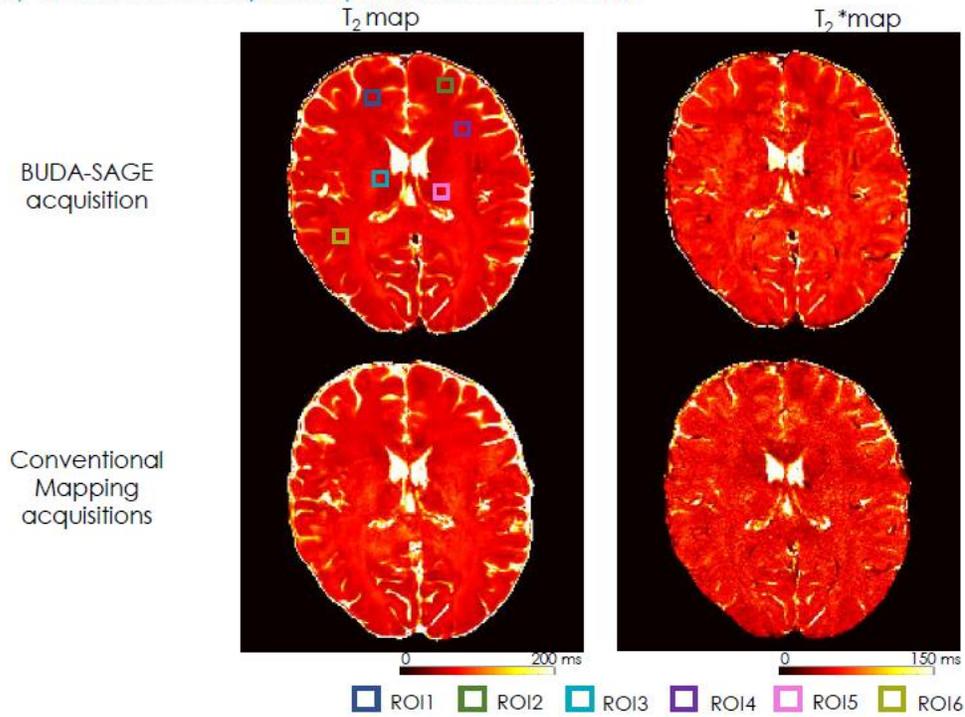

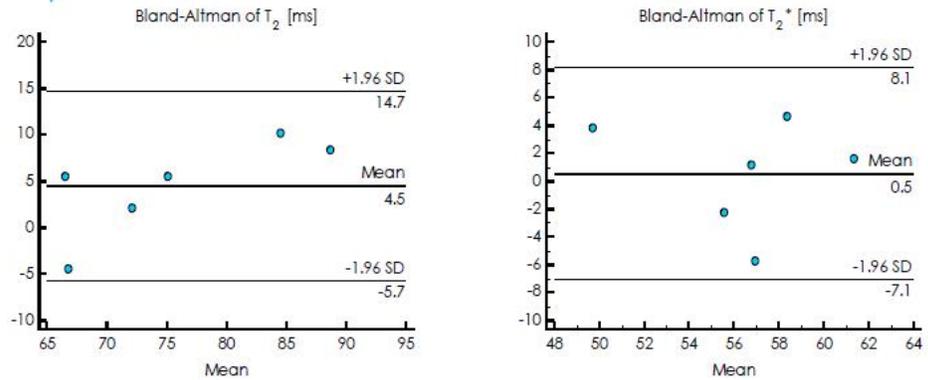

Figure 4

(a) Distortion-free quantitative $T_2$ and $T_2^*$ maps of in-vivo data at $1\times 1\times 2$ mm$^3$ resolution were obtained using Bloch dictionary matching. The maps from a BUDA-SAGE msEPI acquisition (top) are also compared with maps derived from conventional approaches (bottom). Six ROIs from the slice were identified for comparison.

(b) The Bland-Altman plots showing the mean and difference for the ROIs for $T_2$ and $T_2^*$ values. All values are within the limits of agreement.

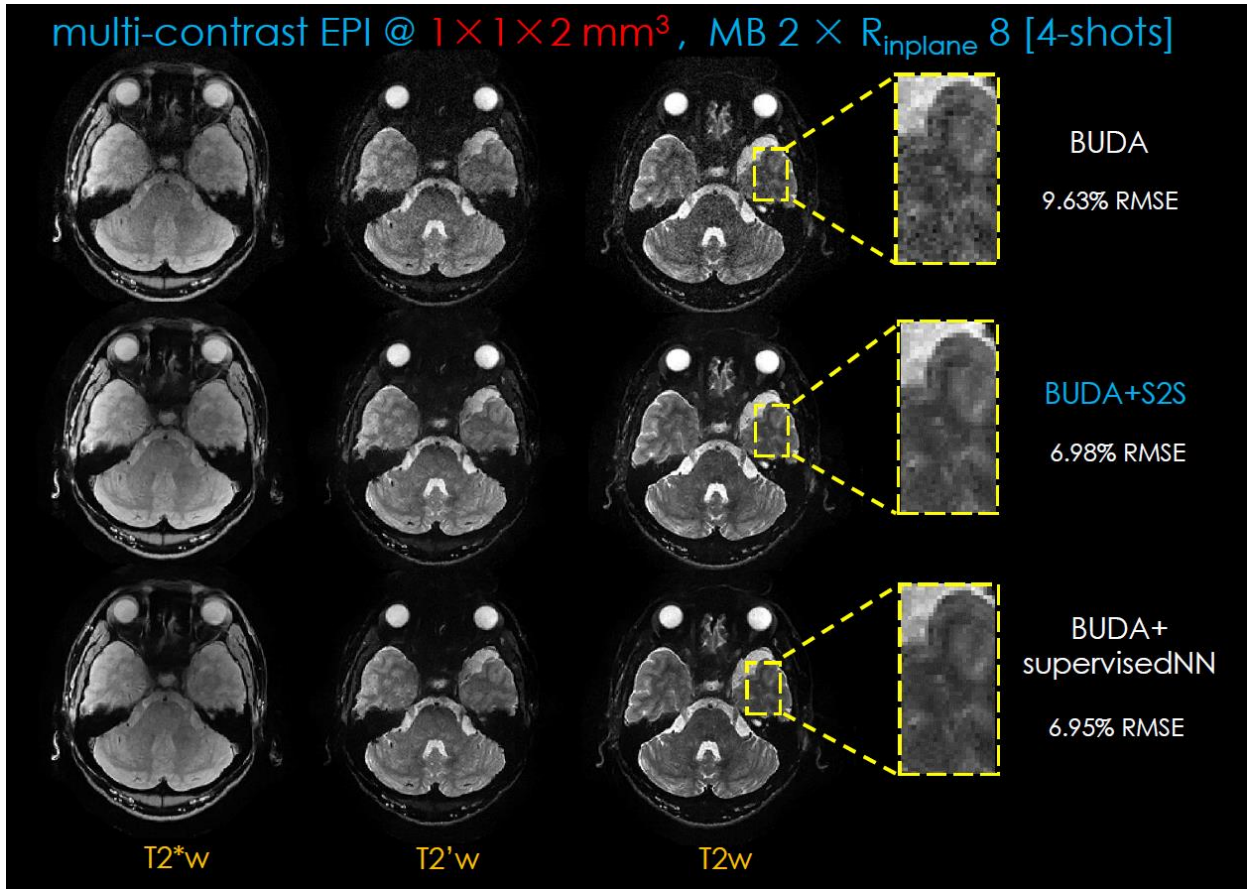

Figure 5

$R_{inplane} \times MB = 8 \times 2$-fold accelerated BUDA-SAGE msEPI acquisition with 4 shots at $1 \times 1 \times 2 mm^3$ resolution. The acquisition time for whole-brain coverage across 60 slices was 25s (20s+5s dummy). The distortion-free, multi-contrast T2*-, T2'- and T2-weighted images were reconstructed by (i) BUDA reconstruction, (ii) BUDA+self-supervised MR-S2S denoising, and (iii) BUDA+supervised neural network denoising.

The RMSE was computed between each method using 4-shot data with respect to a reference 8-shot BUDA reconstruction. Compared to standard BUDA, BUDA+MR-S2S effectively reduced the noise. The results from BUDA+MR-S2S are comparable with those from BUDA+supervisedNN based on the RMSE values and qualitative inspection.

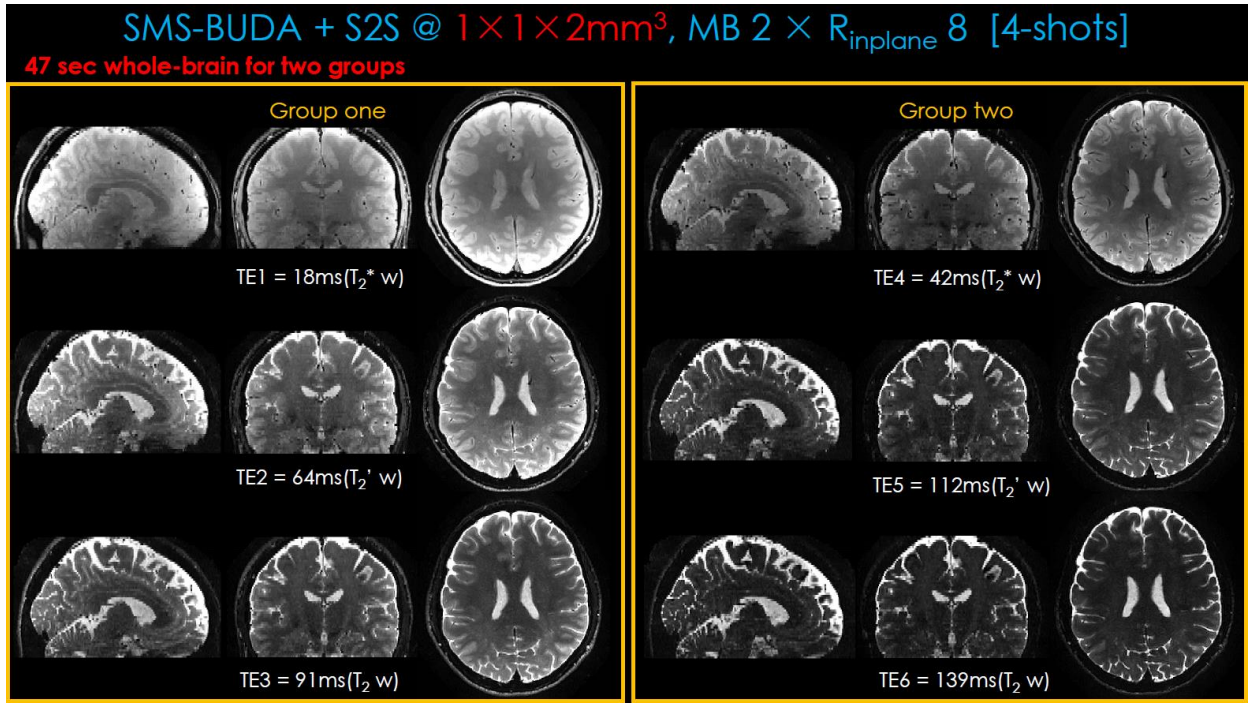

Figure 6

$R_{inplane} \times MB = 8 \times 2$-fold accelerated BUDA-SAGE msEPI acquisition with 4 shots and 2 groups at $1 \times 1 \times 2 mm^3$ resolution. [TEgre, TEmixed, TEse]=[18, 64, 91] ms in group 1 and [42, 112, 139] ms in group 2 at TR=5000 ms. The total acquisition time for whole-brain coverage is 47s (20s/group + 5s dummy), including a 2s calibration scan for coil sensitivities.

Distortion-free six-contrast images (2 groups, each group comprising 3 contrasts) reconstructed by BUDA+self-supervised Self2Self (BUDA+MR-S2S) are depicted.

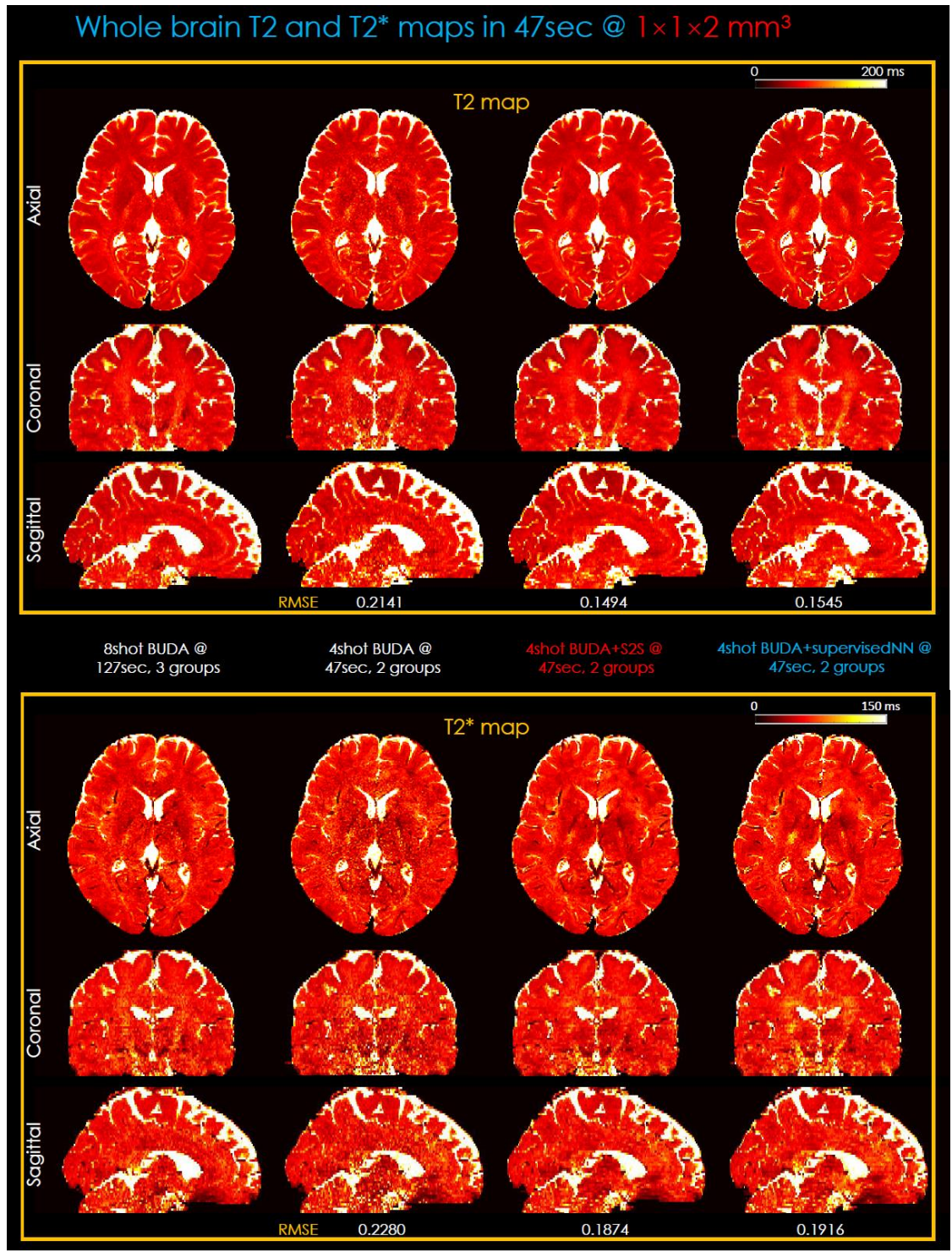

Figure 7

Whole-brain, distortion-free $T_2$ and $T_2^*$ maps at $1\times1\times2$ mm$^3$ resolution were obtained using Bloch dictionary matching from the multi-contrast images. An 8-shot, 3-groups data (acquisition time: 127s = 40s/group + 5s dummy + 2s calibration) was used as reference. The depicted maps are based on the reconstruction methods of BUDA, BUDA+MR-S2S, BUDA+supervisedNN which use 2-groups, 4-shot data

(acquisition time: 47s = 20s/group + 5s dummy + 2s calibration). BUDA+MR-S2S with 2 groups, 4 shots data obtained comparable maps respect to those from BUDA+supervisedNN

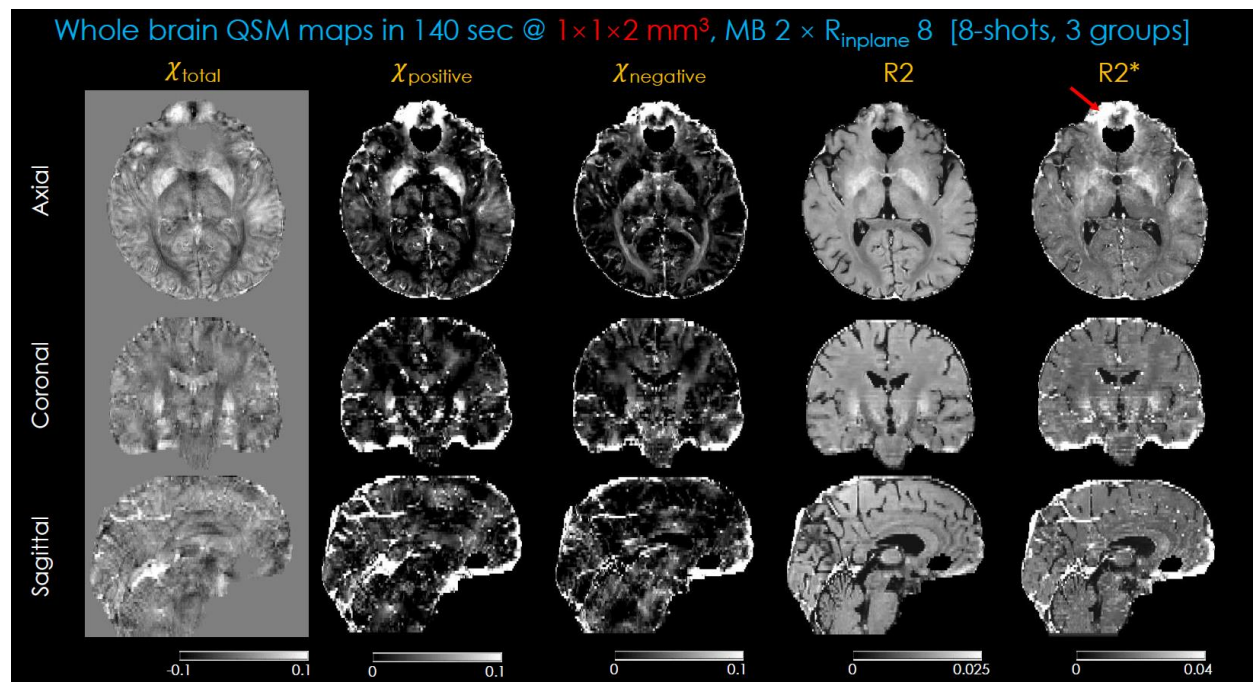

Figure 8

Whole-brain, distortion-free total, para- and dia-magnetic QSM maps along with R2, R2* maps using 3 groups, 8 shots data at $1\times1\times2$ mm$^3$ resolution. Data were acquired using the BUDA-SAGE sequence with $R_{inplane}$=8, SMS=2. The acquisition time for whole-brain coverage across 64 slices was 137.5s (44s/group+5.5s dummy), plus a 2s calibration scan for coil sensitivities.

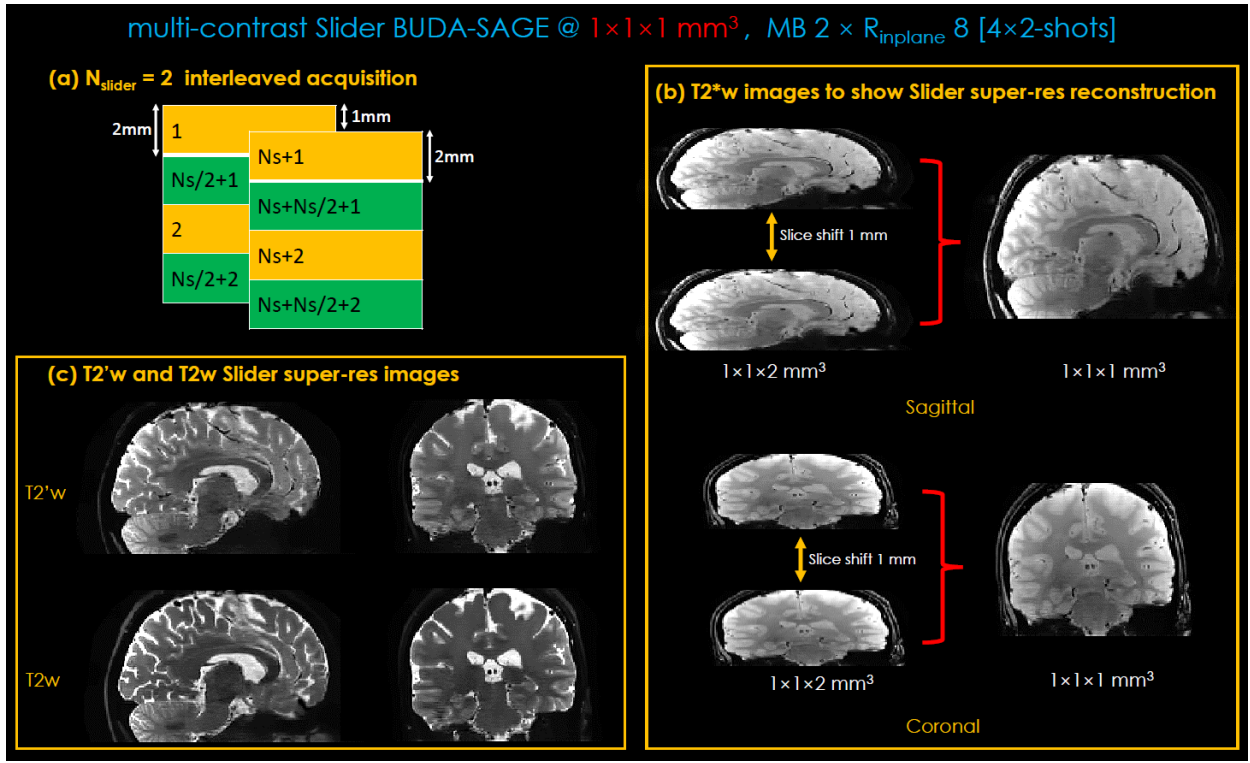

Figure 9

(a) $N_{slider}$ =2 interleaved Slider-encoding acquisition.

(b)&(c) $R_{inplane} \times MB = 8 \times 2$-fold accelerated, Slider-encoded BUDA-SAGE msEPI acquisition, which provided $1 \times 1 \times 1 mm^3$ resolution images after super-resolution reconstruction of two volumes shifted by 1 mm in the slice direction.

The acquisition time for whole-brain coverage was 45s (20s/slider×2+5s dummy). The distortion-free, multi-contrast images including T2*-, T2'- and T2-weighted images were reconstructed using BUDA+MR-S2S.

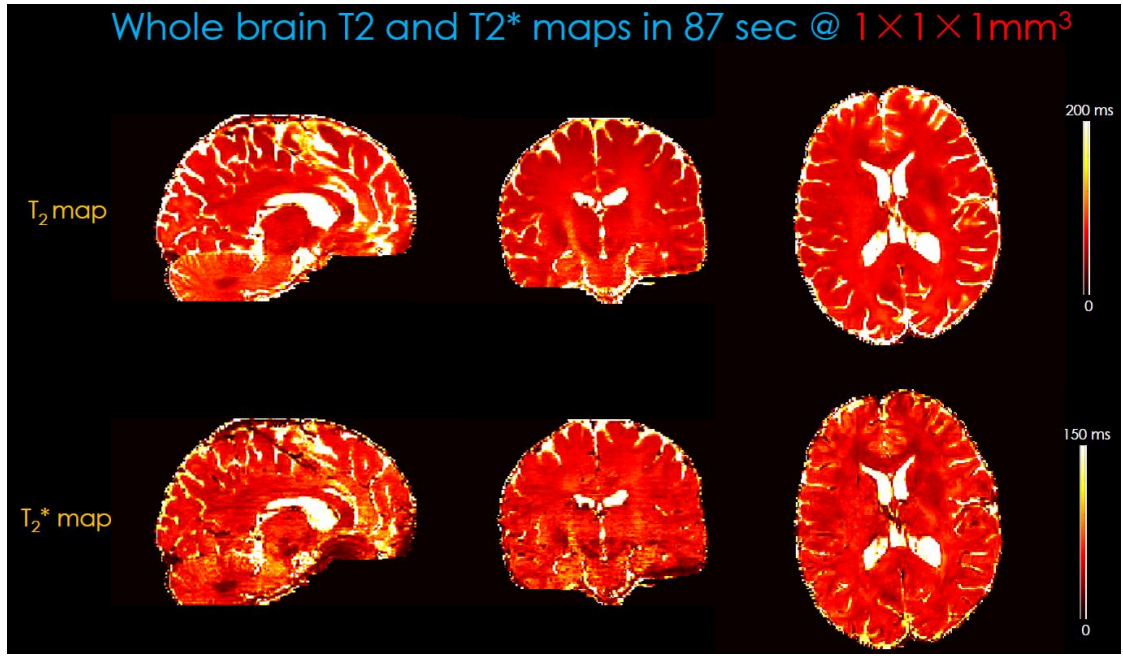

Figure 10

Whole-brain, distortion-free $T_2$ and $T_2$* maps at $1 \times 1 \times 1$ mm³ resolution were obtained using Bloch dictionary matching on the multi-contrast images. These maps are based on BUDA+MR-S2S reconstruction with 2-groups, 4-shot Slider-encoded ($N_{\text{slider}} = 2$) data (acquisition time: 87s = 20s/group/slider + 5s dummy + 2s coil calibration) .

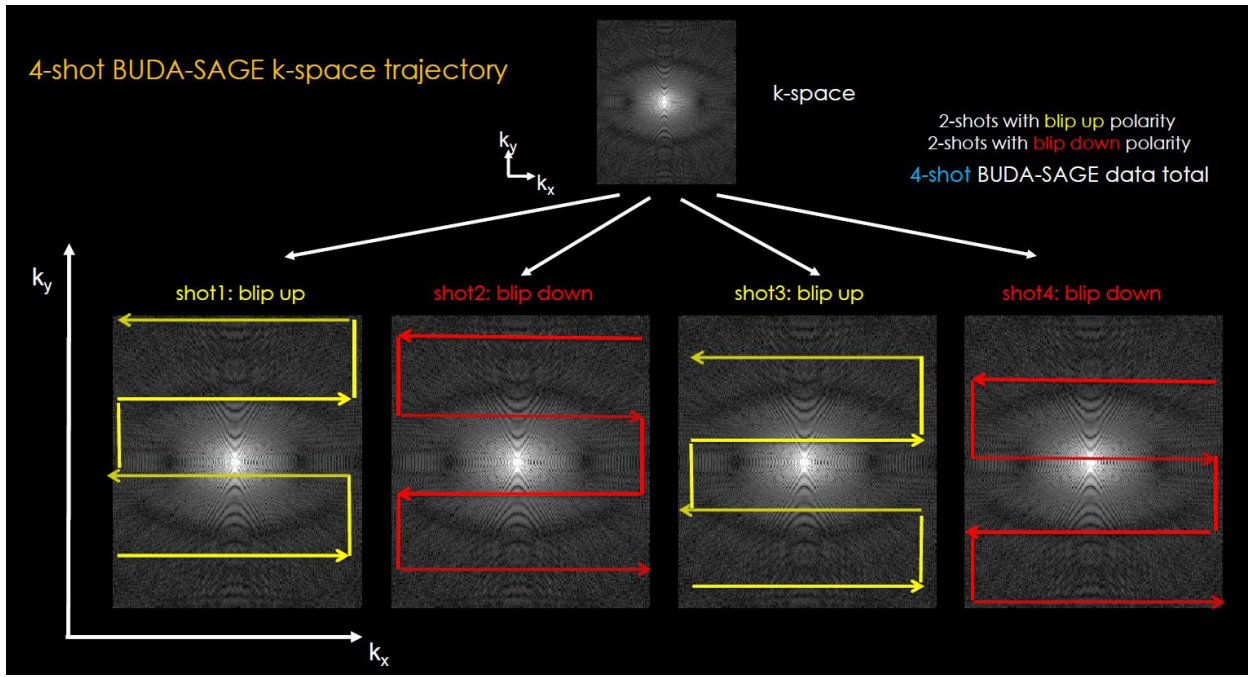

Figure S1

The k-space trajectory of 4-shot BUDA-SAGE EPI acquisition, where successive shots are made with alternating phase encoding polarity.

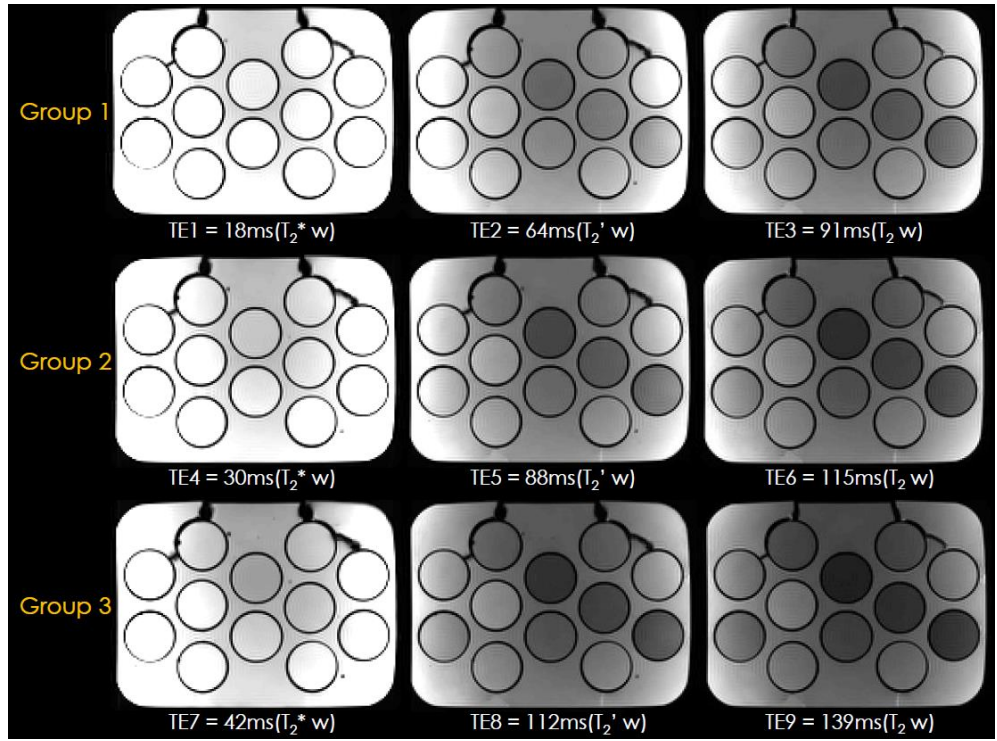

Figure S2

$R_{inplane} \times MB = 8 \times 1$-fold accelerated BUDA-SAGE msEPI acquisition with 8 shots at $1 \times 1 \times 2 mm^3$ resolution on the tube phantom. [TEgre, TEmixed, TEse]=[18, 64, 91] ms in group 1, [TEgre, TEmixed, TEse]=[30, 88, 115] ms in group 2 and [42, 112, 139] ms in group 3. The distortion-free, multi-contrast T2*-, T2'- and T2-weighted images were obtained by BUDA-SAGE reconstruction.

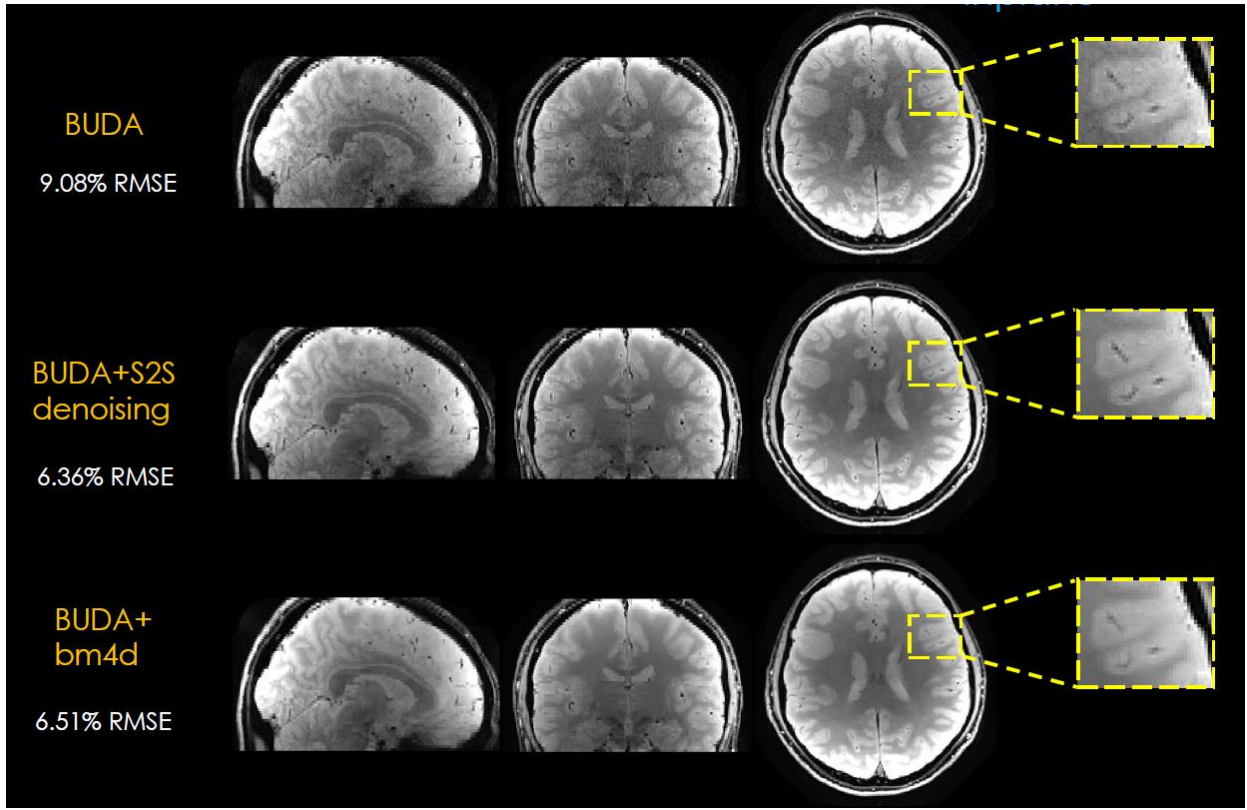

Figure S3

The comparison results are reconstructed by (i) BUDA reconstruction (ii) BUDA + self-supervised MR-S2S denoising (iii) BUDA + BM4D denoising. Compared to standard BUDA, BUDA+MR-S2S effectively reduced the noise. However, the BM4D result looks a bit over-smooth seen from the blurry structure details.

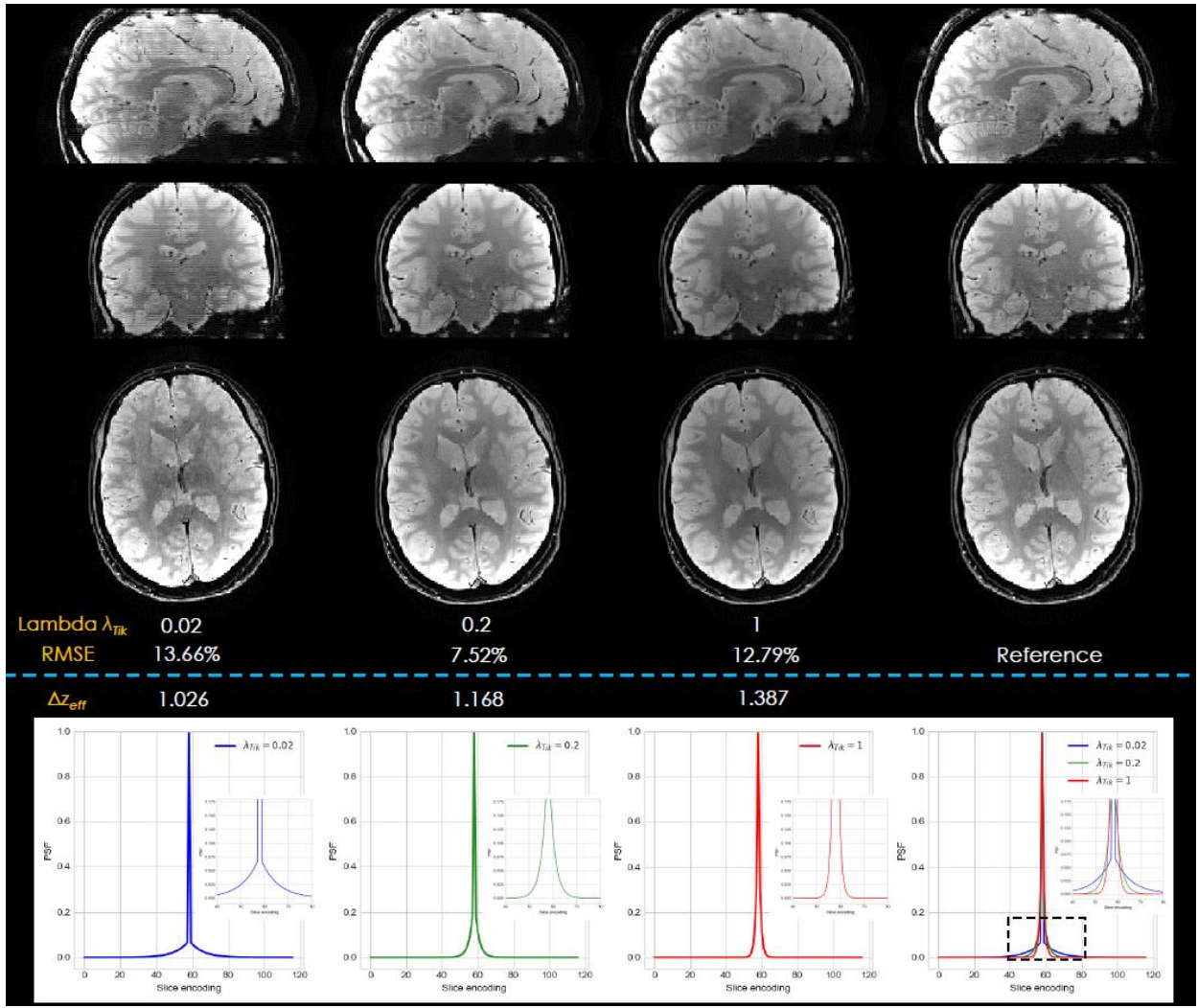

Figure S4

Comparison of $T_2^*$-weighted images at 1×1×~1 mm$^3$ resolution based on different $\lambda_{Tik}$ for Tikhonov regularization from 0.02 to 1. The impulse responses and $\Delta z_{eff}$ were obtained. The reference is from the 8-shot Slider-encoded BUDA-SAGE acquisition.